\documentclass[aps,prb,twocolumn,groupedaddress,showpacs,preprintnumbers,amsmath,amssymb,floatfix]{revtex4}
\usepackage{graphicx}
\usepackage[tight]{subfigure}
\usepackage{dcolumn}
\usepackage{float}
\usepackage{bm}
\usepackage{enumerate}

\newcommand{\PotGeneral}{{0}}
\newcommand{\PotBab}{{1}}
\newcommand{\PotRotBab}{{1_{\rm rot}}}
\newcommand{\PotB}{{2}}
\newcommand{\PotRho}{{\rm eff}}

\renewcommand{\Im}{\mathrm{Im}}
\newcommand{\Ordo}{O}

\begin{document}

\title{Investigation of the stability of Hopfions in the two-component
  Ginzburg-Landau model}

\author{Juha J\"aykk\"a}
\email{juolja@utu.fi}
\author{Jarmo Hietarinta}
\email{hietarin@utu.fi}
\affiliation{
Department of Physics,
University of Turku,
FI-20014 Turku, Finland
}

\author{Petri Salo}
\email{Petri.Salo@tkk.fi}
\affiliation{
Laboratory of Physics, Helsinki University of Technology, P.O.\ Box
1100, FI-02015 TKK, Espoo, Finland
}

\date{\today}

\begin{abstract}
  We study the stability of Hopfions embedded in the Ginzburg-Landau
  (GL) model of two oppositely charged components. It has been shown
  by Babaev {\it et~al.} [Phys.\ Rev.\ B {\bf 65}, 100512 (2002)] that
  this model contains the Faddeev-Skyrme (FS) model, which is known to
  have topologically stable configurations with a given Hopf charge,
  the so-called Hopfions. Hopfions are typically formed from a
  unit-vector field that points to a fixed direction at spatial
  infinity and locally forms a knot with a soft core.  The GL model,
  however, contains extra fields beyond the unit-vector field of the FS
  model and this can in principle change the fate of topologically
  non-trivial configurations.  We investigate the stability of
  Hopfions in the two-component GL model both analytically (scaling)
  and numerically (first order dissipative dynamics). A number of
  initial states with different Hopf charges are studied; we also
  consider various different scalar potentials, including a singular
  one. In all the cases studied, we find that the Hopfions tend to
  shrink into a thin loop that is too close to a singular configuration for our
  numerical methods to investigate.
\end{abstract}

\maketitle

\section{\label{sec:intro}Introduction}

Topologically stable knots and other vortex-like structures have recently received wide interest within
condensed matter physics. One reason for this is the continuous development of experimental techniques which
now allow the production of vortices in various types of media. These include, e.g., Bose-Einstein
condensates~\cite{AboShaeer:2001aa, Engels:2003aa}, superconductors~\cite{Monaco:2005fi, Monaco:2002aa,
  Carmi:2000}, super-fluid $^3$He~\cite{Parts:1995,Finne:2004aa} and nematic liquid crystals~\cite{Ray:2001ug}.
At the same time the increased capacity of supercomputers has made it possible to study these structures
numerically.  Thus, in the last decade there have been many numerical studies devoted to finding stable
topologically non-trivial configurations in different physical systems, including, e.g., topological un-knots,
knots and vortices in the Faddeev-Skyrme (FS) model~\cite{Faddeev:1997zj, Battye:1998pe, Battye:1998zn,
  Hietarinta:1998kt, Hietarinta:2000ci, Hietarinta:2003vn, Adam:2006bw}, vortices in Bose-Einstein
condensates~\cite{Martikainen:2002gf, Mackie:2003dk, Ruostekoski:2005zd}, vortex atom lasers in a two-flavor
Bose condensate~\cite{Liu:2006aa} and vortices in superconductors~\cite{Smorgrav:2004ky, Donaire:2004gp},
liquid helium~\cite{Eltsov:2006aa, Finne:2003aa}, liquid metallic hydrogen~\cite{Babaev:2004rm} and possibly
even in neutron stars~\cite{Babaev:2002wa}.

There are various ways in which a vector field in nature can support vortices. For example a velocity
field can form a vortex, e.g., in a hurricane. The position of the vortex is in the eye of the hurricane,
where the velocity is zero. Such a vortex is not stable, because a velocity field can continuously change to
zero everywhere.

In some special materials there can exist vector fields, associated to spin or other such property, that
cannot vanish.  Then it is possible to have vortices that are both non-singular and conserved, the
conservation following from topological reasons.  (One example that can be produced in a laboratory is the
continuous unlocked vortex (CUV) in super-fluid ${}^3$He-A~\cite{Parts:1995}.)  The prototype model that
contains non-vanishing topological structures characterized by a Hopf charge was presented by L.\
Faddeev~\cite{Faddeev:1976,Faddeev:1997zj}, and recently it has been shown numerically that this model does
indeed contain stable topological solitons~\cite{Battye:1998pe, Battye:1998zn, Hietarinta:1998kt,
Hietarinta:2000ci}.

In this work we study numerically a system of two electro-magnetically coupled, oppositely charged Bose
condensates described by the Ginzburg-Landau (GL) model. Babaev {\it et al.}~studied this system in
Ref.~\cite{Babaev:2001zy} and argued that it should contain stable knotted vortex solitons with a non-zero Hopf
charge.  Here we present our results for the GL model, the main conclusion being that Hopfions in the GL model
seem to be unstable.

The paper is organized as follows: In Section \ref{sec:model} we formulate the equations, discuss the possible
potentials and present the initial states that are used in the computation. In Section \ref{sec:numerics} we
present the discretization of the GL Lagrangian and the numerical method for finding stable minimum energy
configurations. In Section \ref{sec:results} we describe the results and finally in Section
\ref{sec:conclusions} we give some concluding remarks on the results obtained.

\section{\label{sec:model}Model}

\subsection{\label{sec:GLmodel}Ginzburg-Landau model}

The model describes two electro-magnetically coupled, oppositely charged Bose condensates, as given by the GL
Lagrangian density
\begin{eqnarray}
   {\cal L}  &=&
   \tfrac{\hbar^2}{2m_1} \Bigl|\bigl(  \nabla +
        i \tfrac{2e}{\hbar c} \vec{A} \bigr)\Psi_1\Bigr|^2 +
   \tfrac{\hbar^2}{2m_2}\Bigl|\bigl(  \nabla -
        i \tfrac{2e}{\hbar c} \vec{A} \bigr)\Psi_2\Bigr|^2 \nonumber\\
          &&  +V\bigl(\Psi_{1},\Psi_{2}\bigr) +
     \tfrac{1}{2\mu_0}\vec{B}^2,
\label{eq:GL}
\end{eqnarray}
where we have used SI units. In Eq.\ \eqref{eq:GL} $\Psi_1$ and $\Psi_2$ are the order parameters for the
condensates, $\vec A$ is the electromagnetic vector potential, $\vec B$ the magnetic field, $\vec B=\tfrac1c
\nabla \times\vec A$, and $V$ is a potential.

Babaev {\it et al.}~\cite{Babaev:2001zy} introduced new variables by setting
\begin{equation}
\label{eq:newvars} \Psi_\alpha = \sqrt{2 m_\alpha}\, \rho \chi_\alpha,
\end{equation}
where the new complex field $\chi$ is normalized as
\begin{equation}
  \lvert \chi_1 \rvert^2 + \lvert \chi_2 \rvert^2 = 1,\label{eq:chisum}
\end{equation}
and therefore the real field $\rho$ is given by
\begin{align}
  \rho^2 = \frac12 \left( \frac{\lvert \Psi_1\rvert^2}{m_1} +
   \frac{\lvert \Psi_2\rvert ^2}{m_2} \right). \label{eq:defofrho}
\end{align}
In terms of the new fields of \eqref{eq:newvars} Eq.\
\eqref{eq:GL} becomes
\begin{align}
    {\cal L} &=
    \hbar^2\rho^2 \Bigl(
      \left| \left(\nabla + i \tfrac{2e}{\hbar c} \vec{A} \right) \chi_1 \right|^2 +
      \left| \left(\nabla - i \tfrac{2e}{\hbar c} \vec{A} \right) \chi_2 \right|^2
    \Bigr) \notag \\
      & \quad +\hbar^2\bigl(\nabla \rho\bigr)^2 +
      V\bigl(\chi_{1},\chi_{2}, \rho^2\bigr) +\tfrac{1}{2\mu_0}\vec{B}^2.
  \label{eq:mFS}
\end{align}

The Lagrangian \eqref{eq:GL} is invariant under the gauge transformation
\begin{equation}
  \label{eq:gaugetr}
  \begin{cases}
    \Psi_1 &\rightarrow e^{-i\tfrac{2e}{\hbar}\theta(x)}\Psi_1\\
    \Psi_2 &\rightarrow e^{i\tfrac{2e}{\hbar}\theta(x)}\Psi_2\\
    A_\mu &\rightarrow A_\mu + c\partial_\mu \theta(x),
  \end{cases}
\end{equation}
and the corresponding Noether current is
\begin{align}
  J_k :&= \tfrac{i\hbar e}{m_1}\bigl(\Psi_1^*\partial_k\Psi_1
  - \Psi_1\partial_k\Psi_1^*\bigr)
  -\tfrac{i\hbar e}{m_2}\bigl(\Psi_2^*\partial_k\Psi_2
  - \Psi_2\partial_k\Psi_2^*\bigr)\nonumber \\
  &\quad-\tfrac{4e^2}{c}\left(\frac{\lvert \Psi_1 \rvert^2}{m_1} +
  \frac{|\Psi_2|^2}{m_2}\right)A_k,
  \label{eq:isoJ}
  \intertext{which, using the Amp\`ere's law, must satisfy}
  \label{eq:Ampere}
  \vec{J} &= \nabla \times \vec{B} = \tfrac{1}{c}\nabla \times \nabla \times \vec{A}.
\end{align}
When we use the new variables \eqref{eq:newvars} here, we obtain
\begin{eqnarray}
  J_k &= & 2e\hbar \rho^2i\bigl(\chi_1^*\partial_k\chi_1
  - \chi_1\partial_k\chi_1^*
  -\chi_2^*\partial_k\chi_2 + \chi_2\partial_k\chi_2^*\bigr)\nonumber\\
  &&-\tfrac{8e^2\rho^2}{c} A_k   \nonumber \\
  &=& 4e\hbar\rho^2 \left( \tfrac12 j_k - \tfrac{2e}{\hbar
      c}A_k\right),\label{eq:Jbychi}
\end{eqnarray}
which contains a new (non-gauge invariant) current~\cite{comment1}
\begin{equation}
  \label{eq:jdef}
  j_k = i\bigl(\chi_1^*\partial_k\chi_1 - \chi_1\partial_k\chi_1^*
  -\chi_2^*\partial_k\chi_2 + \chi_2\partial_k\chi_2^*\bigr).
\end{equation}
Later on we also use a gauge invariant vector field
\begin{equation}
  \label{eq:Cdef}
  \vec{C} := \tfrac{1}{\hbar e \rho^2}\vec{J}.
\end{equation}

Next we define the unit vector field $\vec n$ by
\begin{equation}\label{eq:defofn}
  \vec{n} = \begin{pmatrix} \chi_1^* & \chi_2 \end{pmatrix}
  \vec{\sigma}
  \begin{pmatrix} \chi_1\\ \chi_2^* \end{pmatrix}
  = \begin{pmatrix}
    \chi_1\chi_2 +\chi_1^*\chi_2^* \\
    i(\chi_1\chi_2 -\chi_1^*\chi_2^*) \\
    |\chi_1|^2 - |\chi_2|^2
  \end{pmatrix},
\end{equation}
where $\vec{\sigma}$ are the Pauli matrices. The inverse transformation is
\begin{equation}\label{eq:n2chi}
  \left\{
    \begin{array}{rcl}
      \chi_1 &=&\frac{n_1-in_2}{\sqrt{2(1-n_3)}}e^{i\alpha}\\
      \chi_2 &=&\frac{\sqrt{(1-n_3)}}{\sqrt 2}e^{-i\alpha},\\
    \end{array}
  \right.
\end{equation}
where the phase must be chosen so that $\chi_\alpha$ are continuous,
see, e.g., \eqref{eq:identification}.
(This inverse transformation is not used in the numerical
simulations.)

In order to show the similarities between the GL and FS models, we write the Lagrangian \eqref{eq:mFS} in
terms of $\vec n,\rho$ and $\vec C$, after which the $\vec n$ part of the result should be similar to the FS
model, given by
\begin{equation}
 \mathcal{L}_{\rm FS}=\tfrac12
\partial_k n_l \partial^k n^l+g_{\rm FS}(\vec{n} \cdot \partial_k
\vec{n} \times \partial_l \vec{n})^2.
\label{eq:trueFS}
\end{equation}

After inverting Eqs.\ \eqref{eq:Jbychi} and \eqref{eq:Cdef} to obtain $\vec{A}$ in terms of $\vec{C}$ and
$\vec{j}$ the kinetic part of Eq.\ \eqref{eq:mFS} becomes
\begin{align}
  \begin{split}
  \mathcal{L}_{\rm kinetic} =\;
  &\hbar^2\rho^2\bigl(
  |\nabla \chi_1|^2 + |\nabla \chi_2|^2 - \tfrac{1}{4}\vec{j}^2
  \bigr) \\
  & + \hbar^2 \bigl(\nabla \rho\bigr)^2 +
  \tfrac{\hbar^2\rho^2}{16}\vec{C}^2.
\label{eq:kinFS}
  \end{split}
\end{align}
Using Eqs.\ \eqref{eq:defofn}, \eqref{eq:chisum} and \eqref{eq:jdef} one finds that
\begin{equation}
  \label{eq:B6}
  \partial_k n_l \partial^k n^l =
  4\bigl(|\nabla \chi_1|^2 + |\nabla \chi_2|^2\bigr) - \vec{j}^2
\end{equation}
and therefore the first term in Eq.\ \eqref{eq:kinFS} corresponds to the first term in Eq.\ \eqref{eq:trueFS}.

By direct substitution of Eqs.\ \eqref{eq:Jbychi} and \eqref{eq:Cdef} into $\vec{B}$, we also find that
\begin{equation}
  \vec{B} = \tfrac{1}{c} \nabla \times \vec{A} =
  \tfrac{\hbar}{4e} \bigl( \nabla \times \vec{j} - \tfrac{1}{2}
  \nabla \times \vec{C} \bigr),\label{eq:BbyCj}
\end{equation}
and again using Eqs.\ \eqref{eq:defofn}, \eqref{eq:chisum} and
\eqref{eq:jdef}, we get
\begin{equation}
  \label{eq:rotjii}
  \tfrac{1}{2} \epsilon_{klm} \vec{n} \cdot \partial_l \vec{n} \times \partial_m \vec{n} =
  -\epsilon_{klm}\partial_l j_m,
\end{equation}
from which we can see that the $\vec B^2$ term in Eq.\ \eqref{eq:mFS} contributes to the second term in Eq.\
\eqref{eq:trueFS}.

Combining the above results we can write the Lagrangian \eqref{eq:GL} in the form
\begin{align}\label{eq:GLfinal}
  \begin{split}
    \mathcal{L} =&\tfrac{\hbar^2 \rho^2}{4} \partial_k n_l \partial^k n^l
    + \hbar^2 \bigl(\nabla \rho\bigr)^2 + \tfrac{\hbar^2\rho^2}{16} \vec{C}^2
    + V\bigl(\rho, n_k\bigr) \\
    &  + \tfrac{\hbar^2}{128\mu_0 e^2} \left[ \epsilon_{klm}
    \bigl( \vec{n} \cdot \partial_k \vec{n} \times \partial_l \vec{n}
    + \partial_k C_l \bigr) \right]^2,
  \end{split}
\end{align}
which is the form derived by Babaev {\it et al.}~\cite{Babaev:2001zy}. The dynamical fields are now $\rho$,
$\vec n$ and $\vec C$. If $\rho=$ constant and $\vec C=0$, the GL model reduces to the FS model in Eq.\
\eqref{eq:trueFS}. Since the FS model contains stable topological structures with non-trivial Hopf charge, one
can hope that the GL model also contains similar structures. However, GL contains the additional
fields $\rho$ and $\vec C$ in comparison to FS and the role of these new fields must be investigated.

\subsection{\label{subsec:potential}Form of the potential}

A typical and rather general quartic potential used in the GL model is
\begin{eqnarray}
  V_\PotGeneral\left(\Psi_1, \Psi_2\right) &=& \tfrac12 c_1 |\Psi_1|^4 +
  \tfrac12 c_2 |\Psi_2|^4 +c_3 |\Psi_1|^2|\Psi_2|^2\nonumber\\
&&+ b_1 |\Psi_1|^2 + b_2 |\Psi_2|^2 + a_0.\label{eq:GenPot}
\end{eqnarray}
When $\Psi_\alpha$ are expressed in terms of $\rho$ and $\vec n$ and the whole system is rescaled so that $m_\alpha
\rightarrow 1$, we find
\begin{equation}\label{eq:n3andpsi}
 |\Psi_1|^2 = \rho^2\bigl(1+n_3\bigr),\quad
 |\Psi_2|^2 = \rho^2\bigl(1-n_3\bigr),
\end{equation}
and then the potential \eqref{eq:GenPot} reads
\begin{eqnarray}
V_\PotGeneral\left(\rho^2, n_3\right)&=&
\rho^4[n_3^2(\tfrac12(c_1+c_2)-c_3)+n_3(c_1-c_2)\nonumber \\ &&
\hspace{1cm} +\tfrac12(c_1+c_2)+c_3]\nonumber \\
&&+\rho^2[n_3(b_1-b_2)+b_1+b_2]+a_0.\label{eq:GenPotN}
\end{eqnarray}

One important aspect in choosing the potential is that at infinity the fields will settle to the minimum of
the potential. Furthermore, in order to define the Hopf charge it is necessary that the $\vec n$ field points
to the same direction far away, otherwise we cannot compactify the 3D-space. It would therefore be optimal to
have a potential with a minimum that would fix the $\vec n$ field completely, say, to $n_3=1$.

For a particular example assume that $c:=c_1=c_2=c_3>0,\,b:=b_1=b_2$, then $n_3$ disappears from Eq.\
\eqref{eq:GenPotN} and the potential minimum is at $\rho^2=-b/(2c)$, $\vec n$ being free. For more generic
parameter values the extrema are obtained for particular values of $n_3$ and $\rho^2$:
\begin{align}
n_3=&\frac{b_1(c_2+c_3)-b_2(c_1+c_3)}{b_1(c_2-c_3)+b_2(c_1-c_3)},\\
\rho^2=&-\frac{b_1(c_2-c_3)+b_2(c_1-c_3)}{2(c_1c_2-c_3^2)}.
\end{align}
This is a minimum, if $c_1c_2>c_3^2$. Note that the above values do not necessarily fall within the allowed
values for $\rho^2$ and $n_3$, (i.e., $\rho^2 > 0$ and $|n_3| \le 1$), in which case the extrema are on the
boundaries of the allowed values.

From physical arguments the following special case is relevant~\cite{Babaev:2002pri}
\begin{equation}\label{eq:BabPotPsi}
  V_\PotBab \left(\Psi_1, \Psi_2\right)   = \lambda \Bigl(
  \bigl( |\Psi_1|^2 - 1 \bigr)^2 +
  \bigl( |\Psi_2|^2 - 1 \bigr)^2 \Bigr).
\end{equation}
This breaks $O(3)$ to $O(2)$ and corresponds to two independently conserved condensates. It has a minimum at
$n_3=0$, $\rho^2=1$.

Another physically relevant~\cite{Babaev:2002pri} potential is
\begin{eqnarray}\label{eq:BabCurPot}
  V_\PotB \left(\Psi_1, \Psi_2\right)   &=& \lambda \Bigl(
  \bigl( |\Psi_1|^2 - 1 \bigr)^2 +
  \bigl( |\Psi_2|^2 - 1 \bigr)^2\Bigr)\nonumber\\
  && +  c\bigl|\Psi_1\Psi_2^* - \Psi_2\Psi_1^*\bigr|+a_0,
\end{eqnarray}
which breaks $O(3)$ completely. This corresponds to multi-band superconductors with inter-band Josephson
effect~\cite{Babaev:2002pri}.  In terms of $\vec n$ this potential is given (using Eq.\ \eqref{eq:n2chi})
\begin{eqnarray}
V_\PotB&=&\lambda\tfrac12n_3^2+c\Im[(n_2-in_1)e^{i2\alpha}]+b_0.
\end{eqnarray}
The minimum of this potential is located at $\rho^2=1/2,\, n_3=\sqrt{1-c^2/\lambda^2}$, while the specific
values of $n_1$ and $n_2$ also depend on $\alpha$, which is related to the phase of $\Psi$.

From the point of view of Hopf-charge conservation, the possibility of $\rho=0$ is problematic, even if it
happens locally, since then the field $\vec n$ is not defined. However, it has been
argued~\cite{Faddeev:2000rp} that quantum effects ensure that $\rho \ne 0$ everywhere.  In classical field
theories, like the present one, such expected quantum effects can be included through effective potentials.
Since the main purpose of this effective potential is to guarantee that $\rho \ne 0$, its exact form is not so
important. One such potential, which we will use later, is
\begin{align}
  \label{eq:effectivepot}
  \begin{split}
  V_\PotRho\left(\Psi_1, \Psi_2\right) =\;&\tfrac{1}{4} \lambda
  \bigl( |\Psi_1|^2 + |\Psi_2|^2 - \rho_0^2 \bigr)^2\\
  &+2\gamma \bigl(|\Psi_1|^2 + |\Psi_2|^2\bigl)^{-1} + a_0.
  \end{split}
\end{align}
The constants $\rho_0$ and $a_0$ are determined by requiring that the minimum of the potential is at
$\rho^2=1$ with a value of 0, yielding $ \rho_0^2 = 2-\gamma/\lambda$ and $ a_0 = - \gamma -
\tfrac{\gamma^2}{4 \lambda}$.

\subsection{\label{subsec:initial_states}Initial states of Hopf invariant Q}
We are interested in the minimum energy configurations of topologically distinct configurations of the
field $\chi$, or by Hopf-map, $\vec{n}$, related to the complex physical fields $\Psi_\alpha$
through Eqs.\ \eqref{eq:newvars} and \eqref{eq:n2chi}.  From the point of view of $\vec n$ it is only
necessary that $\lim_{|\vec x|\to\infty}\vec n=\vec n_\infty$ is the same in all directions. From this it
follows that we can compactify $ \mathbb{R}^3\to \mathrm{S}^3$ and then $\vec{n}$ becomes a map $\mathrm{S}^3
\rightarrow \mathrm{S}^2$ with homotopy group $\pi_3\left(\mathrm{S}^2\right)=\mathbb{Z}$, characterized by
the Hopf charge.

Therefore, we have to create a configuration with $\Psi: \mathbb{R}^3 \rightarrow \mathbb{C}^2$ and
$\vec{A}: \mathbb{R}^3 \rightarrow \mathbb{R}^3, $ such that $\vec{n}$ has the desired property mentioned
above.  For $\Psi_\alpha$ this implies $|\Psi_1|^2 + |\Psi_2|^2 \ne 0$ everywhere and
$\lim_{|\vec{x}|\rightarrow \infty} \Psi_\alpha$ are (independent) constants. The first condition is also
necessary for defining the field $\chi$ and if the the second condition is also satisfied $\chi$ becomes a map
$\mathrm{S}^3 \rightarrow \mathrm{S}^3$.

The method of constructing a configuration with a desired Hopf charge has been investigated by Aratyn {\it et
  al.}~\cite{Aratyn:1999cf}. They used the fact that for any function $\phi:\mathrm{S}^3 \rightarrow
\mathrm{S}^3$ combined with the Hopf map $h:\mathrm{S}^3 \rightarrow \mathrm{S}^2$
\begin{align}
  \label{eq:hopfmap}
  h\left(\phi_1,\phi_2,\phi_3,\phi_4\right) &=
  \begin{pmatrix}
    2(\phi_1 \phi_3 - \phi_2 \phi_4)\\
    -2(\phi_1 \phi_4 + \phi_2 \phi_3)\\
    \phi_1^2 + \phi_2^2 - \phi_3^2 - \phi_4^2
  \end{pmatrix},
\end{align}
the Hopf charge of $h \circ \phi:\mathrm{S}^3 \rightarrow \mathrm{S}^2$ equals the degree of the map
$\phi:\mathrm{S}^3 \rightarrow \mathrm{S}^3$ (the degree is the $n$-dimensional generalization of the
1-dimensional degree, also known as the winding number).
 
For an explicit construction one uses the toroidal coordinates
$\bigl(\eta, \xi, \varphi\bigr)$ of $\mathbb{R}^3$ defined by
\begin{gather}
  \begin{split}
  \begin{aligned}\label{eq:toroidal_coordinates}
    x_1 &=\frac{\sinh(\eta)\cos(\varphi)}{\Delta},
    &\quad x_2&=\frac{\sinh(\eta)\sin(\varphi)}{\Delta},\\
    x_3 &=\frac{\sin(\xi)}{\Delta}, &\quad \Delta&=\cosh(\eta)-\cos(\xi).
  \end{aligned}
  \end{split}
\end{gather}
In these coordinates the core is at $\eta=\infty$, while the $z$-axis and
spatial infinity are at $\eta=0$.

Next take any monotonic function $g\colon[0,\infty\bigl)\, \rightarrow
[0,1]$, (or $[-1,0]$) choose $p,q\in\mathbb{Z}$ and define the map
$\phi\colon\mathrm{S}^3 \rightarrow \mathrm{S}^3$ by
\begin{align}
  \label{eq:AFZphi}
  \begin{split}
    \phi = \bigl( &g(\eta) \cos(p\xi), g(\eta) \sin(p\xi), \\
    &\sqrt{1-g\bigl(\eta\bigr)^2} \cos(q\varphi),
    \sqrt{1-g\bigl(\eta\bigr)^2} \sin(q\varphi)\bigl).
  \end{split}
\end{align} 
If furthermore $g$ is such that $g^2(\infty) - g^2(0) = \pm 1$, then the
combined map, $h \circ \phi$, has the Hopf invariant
\begin{equation}
  \label{eq:Hofhcombmapphi}
  H(h \circ \phi) = \pm pq.
\end{equation}
For details, see Ref.~\cite{Aratyn:1999cf}.  Using \eqref{eq:AFZphi}
we now identify
\begin{subequations}
  \label{eq:identification}
  \begin{align}
    \chi_1:=\phi_1 + i \phi_2 &= g(\eta)e^{ip\xi},\\
    \chi_2:=\phi_3 + i \phi_4 &=\sqrt{1-g(\eta)^2}\,e^{iq\varphi}.
  \end{align}
\end{subequations}
At the $z$-axis $\chi_2$ looks like a $\varphi$-vortex, and therefore
for continuity we add the further requirement that $g(0)=\pm1$.
Similarly, around the core (located at $\eta=\infty$) we have a
$\xi$-vortex, and therefore we demand that $g(\infty)=0$.

From Eqs.\ \eqref{eq:defofn} and \eqref{eq:identification} we obtain
\begin{align}\label{eq:torus_configuration}
  \vec{n}\bigl(\vec{x}\bigr) &=
  \begin{pmatrix}
    2g\left(\eta\right)\sqrt{1-g^2}\cos(p\xi + q\varphi)\\
    -2g\left(\eta\right)\sqrt{1-g^2}\sin(p\xi + q\varphi)\\
    2g^2\left(\eta\right)-1
  \end{pmatrix}\text{.}
\end{align}
Finally, to close the loop, the $\chi$ of \eqref{eq:identification}
can be recovered from \eqref{eq:n2chi} using
\eqref{eq:torus_configuration} and choosing $\alpha=-q\varphi$.

Note that since $\eta=0$ at infinity and $g(0)^2=1$ the above
construction implies $\vec{n}_\infty :=
\lim_{|\vec{x}|\rightarrow\infty}\vec{n}=(0,0,1)$.

Inverting the toroidal coordinates defined in Eq.\
\eqref{eq:toroidal_coordinates} enables us to express Eq.\
\eqref{eq:identification} in the Cartesian coordinates. In addition,
denoting $r^2=x_1^2+x_2^2+x_3^2$ and choosing $\rho=1$, 
$g(\eta)=1/\cosh(\eta)$ we obtain
\begin{subequations}\label{eq:Psiwithsech}
  \begin{align}
    \Psi_1\bigl(\vec{x}\bigr) &= 
\tfrac{\sqrt{(r^2-1)^2+4x_3^2}}{r^2+1}
\Bigl(\tfrac{r^2-1-2ix_3}{\sqrt{(r^2-1)^2+4x_3^2}}\Bigr)^p, 
\\
    \Psi_2\bigl(\vec{x}\bigr) &=
    \tfrac{2\sqrt{x_1^2+x_2^2}}{r^2+1}
    \Bigl(\tfrac{x_1+ix_2}{\sqrt{x_1^2+x_2^2}}\Bigr)^q.
  \end{align}
\end{subequations}
This is the formula used for the initial configurations of
$\Psi_\alpha$ for our numerical computations. The state indeed has the
correct Hopf invariant, which, in the case of $p=q=1$ can be seen by
finding the preimages of $\Psi_\alpha=0$, which correspond to
preimages of $n_3=\pm 1$. These form two closed loops, namely
$r=1,x_3=0$ for $\Psi_1 = 0$ and the $z$-axis, which in the
compactified space $S^3$ is actually closed.

Fixing $\vec{n}$ or $\chi_i$ does not say anything about the magnitude
of $\Psi$. From energy consideration we must choose $\lim_{|\vec{x}|
  \rightarrow \infty} \Psi_\alpha$ to be one of the minima of the
potential and this fixes a \emph{preferred} value for $\rho$. We use
this preferred value for all $\vec{x}$ when constructing the initial
configurations.

\section{\label{sec:numerics}Numerics}

In this section we will describe the discretization of the model and the method of the minimization of the
Lagrangian \eqref{eq:GL}. We will also compare the GL to FS model and present some test calculations and the
parameters as well as coupling constants for the simulations.

It is important to note that all our simulations are
done with the fields $\Psi_\alpha$ only, the fields $\vec{n}$, $\rho$
and $\vec{C}$ are never used in any computation, they are only used in
the analysis of the results.

It is perhaps useful to mention once again that the change of variables
in Eqs.\ \eqref{eq:newvars}--\eqref{eq:defofrho} is not reversible
whenever $\rho(\vec{x}) = 0$.  There is, however, no guarantee that
$\rho$ stays non-zero in numerical simulations of the physical fields
$\Psi_\alpha$ and $\vec{A}$, even if the minimum of the potential is
at a non-zero value of $\rho$. The situation when $\rho=0$ locally can
imply breakdown in topology and therefore in our numerical simulations
we have monitored the changes in the global minimum value of $\rho$

Topology can also break if the topological structure shrinks smaller
than the lattice unit length. In order to be aware of this possibility
we have monitored the global minimum of the dot product of
nearest-neighbor $\vec n$-vectors: when the global minimum becomes
negative the lattice is probably too coarse, or the configuration has
a genuine singularity. When this has happened in the simulations, we have repeated the simulation with
ever increasing lattice size until the computing resources were exhausted. This strongly points to a singular
configuration.

For simplicity we have used the rescaled $\bigr(|\Psi_\alpha|^2/m_\alpha \rightarrow |\Psi_\alpha|^2\bigr)$
Lagrangian and natural units ($c=\hbar=1$) throughout our numerical work. With these choices, the Lagrangian
density used in all our numerical simulations becomes
\begin{align}\label{eq:cleanL}
  \begin{split}
    {\cal L} &= \tfrac{1}{2} \Bigl|\bigl( \nabla + i g \vec{A} \bigr) \Psi_1\Bigr|^2
    + \tfrac{1}{2}\Bigl|\bigl( \nabla - i g \vec{A} \bigr)\Psi_2\Bigr|^2 \\
    &\quad+ \tfrac{1}{2} g_f (\nabla \times \vec{A})^2 + V\bigl(\Psi_{1,2}\bigr).
  \end{split}
\end{align}
This can be further scaled by $\vec{A} \rightarrow \tfrac{1}{g} \vec{A}$, which reveals the fact that the only
relevant parameter is $g^2/g_f = 4 \mu_0 e^2$. The values used in numerical simulations are $g=1$ and $g_f \in
\{0.01, 1, 2, 100\}$; these amount to a particular choices of units for $\mu_0$ and $e$.

In our studies of the FS model, we have always assumed that $\lim_{|x| \rightarrow \infty} n_3 = 1$, but some
of the present potentials, e.g., $V_\PotBab$, do not have that as a minimum. In this case we may assume that
$\lim_{|x| \rightarrow \infty} n_2=1$, but this can be transformed to $n_3=1$ by a global rotation in the
configuration space, which takes $n_3 \rightarrow n_2$ and $n_2 \rightarrow -n_3$.  The same effect can be
achieved by using new fields defined by
\begin{align}
  \Psi_1' &=\tfrac1{\sqrt 2}(\Psi_1+i\Psi_2^*),\,
  \Psi_2'=\tfrac1{\sqrt 2}(\Psi_1-i\Psi_2^*),
  \label{eq:newpsi}
  \intertext{changing $V_\PotBab$ of \eqref{eq:BabPotPsi} to}
  V_\PotRotBab &= \lambda \bigl(\left(|\Psi_1|^2 + |\Psi_2|^2 - 2\right)^2 \nonumber \\
  &- \left(\Psi_1 \Psi_2 - \Psi_1^* \Psi_2^*\right)^2\bigr).
\end{align}

The different potential terms also contain various parameters. For $V_\PotGeneral$, we have always used
$4c_1=4c_2=2c_3=-b_1=-b_2=a_0$, which, denoting $\lambda \equiv \tfrac{1}{2} c_1$, enables us to write
$V_\PotGeneral = \lambda\bigr(|\Psi_1|^2+|\Psi_2|^2-2\bigl)^2$. In the numerical simulations, we have used
potentials $V_\PotGeneral$, $V_\PotRotBab$ and $V_\PotRho$ with $\lambda \in \{0, 1, 4, 100, 1000\}$.

\subsection{Discretization}
The system has been discretized on a cubic rectangular lattice (indexed as $(s,u,v)$) with periodic boundary
conditions. Our model can be considered as a two-component version of the time-independent, ordinary Abelian
$\mathbb{U}(1)$ Higgs model, which has long since been discretized for lattice simulations in quantum field
theory (for example, see Ref.~\cite{Wilson:1974sk,Damgaard:1988ec} and the references therein).  The main
point in that context is to discretize the fields so that gauge invariance is preserved. From Eq.\
\eqref{eq:gaugetr} we see that if we use the forward discretization of the derivatives, the gauge
transformation of $A_k$ has to be discretized as follows:
\begin{subequations}
 \label{eq:diskrgauge}
 \begin{align}
  A_{1|s,u,v}&\to A_{1|s,u,v}+\tfrac{c}{a}(\theta_{s+1,u,v}-\theta_{s,u,v}),\\
  A_{2|s,u,v}&\to A_{2|s,u,v}+\tfrac{c}{a}(\theta_{s,u+1,v}-\theta_{s,u,v}),\\
  A_{3|s,u,v}&\to A_{3|s,u,v}+\tfrac{c}{a}(\theta_{s,u,v+1}-\theta_{s,u,v}),
 \end{align}
\end{subequations}
where $a$ is the lattice parameter. Thus, $A_k$ should be considered as living on the link between two lattice
points parallel to the coordinate axis $k$. Combining this discretization with the way the $\Psi_\alpha$
transforms in Eq.\ \eqref{eq:gaugetr} it follows that the following combination
\begin{align}
  \Psi&^*_{1|s+1,u,v}\Psi_{1|s,u,v}e^{-i a\kappa A_{1|s,u,v}} \nonumber \\
  +&\Psi_{1|s+1,u,v}\Psi^*_{1|s,u,v}e^{i a\kappa A_{1|s,u,v}} \nonumber \\
  -&\Psi^*_{1|s+1,u,v}\Psi_{1|s+1,u,v} -\Psi^*_{1|s,u,v}\Psi_{1|s,u,v},
 \label{eq:diskrgauge2}
\end{align}
where $\kappa=\frac{2e}{\hbar c}$, is gauge invariant. If we calculate its continuum limit as $a\to 0$ by
expanding in $a$ (e.g., $\Psi_{1|s+1,u,v}= \Psi_1(x+a,y,z)=\Psi_1(x,y,z)+ a\partial_x\Psi_1(x,y,z) +\tfrac12
a^2\partial_x^2\Psi_1(x,y,z)+\dots$) we obtain $-a^2|(\partial_1+i\kappa A_1)\Psi_1|^2+\Ordo(a^3)$.  For
$\Psi_2$ with $A_1$ we use a similar expression with $\kappa\to-\kappa$. Finally, for the other components of
$\vec{A}$ we use corresponding shifts, as illustrated in Eqs.\ \eqref{eq:diskrgauge} and
\eqref{eq:diskrgauge2}.

For the discretization of $\vec B^2$ we use the expression
\begin{align}
  \begin{split}
    \label{eq:Bdiskr}
    e^{iF_{12|suv}}+e^{-iF_{12|suv}}
    &+e^{iF_{23|suv}}+e^{-iF_{23|suv}}\\
    &+e^{iF_{31|suv}}+e^{-iF_{31|suv}}-6,
  \end{split}
  \intertext{where, for example,}
  \begin{split}
    \label{eq:Fdiskr}
    F_{12|suv} = &A_{1,s,u+1,v}-A_{1,s,u,v} \\
    -&A_{2,s+1,u,v}+A_{2,s,u,v}.
  \end{split}
\end{align}
The $F_{kl}$ are gauge invariant under Eq.\ \eqref{eq:diskrgauge}, and the continuum limit of Eq.\
\eqref{eq:Bdiskr} with Eq.\ \eqref{eq:Fdiskr} is $-a^2(\nabla\times \vec A)^2 + \Ordo(a^3)$.

For the potential we use $a^2V(\Psi_{\alpha|suv})$. The discretized Lagrangian is the sum of all the above
terms multiplied by $-a/2$, and then the resulting sum has the continuum limit of Eq.\ \eqref{eq:cleanL}. In
practice we use the cubic lattices of sizes of $60^3$ \ldots $960^3$.  Of these just the sizes $240^3$, $480^3$
and $720^3$ are used in actual simulations, the remaining sizes used only to verify the code, results and
discretization.

The minimization of the Lagrangian has been done using the steepest descent method and the Fletcher-Reeves
variant of the conjugate gradient method. The gradients needed have been calculated symbolically from the
discretized Lagrangian.

\subsection{Initial states}

In order to start the simulations we have to generate initial states with specified Hopf invariants. These
initial configurations have been made using Eq.\ \eqref{eq:Psiwithsech} and choosing $\rho$ such that
$V(\Psi_\infty)=0$. Three different values of Hopf invariant were used: 1, 2 and 4. However, this leaves
$\vec{A}$ undetermined. The topology of the system is unaffected by $\vec{A}$, so in principle, we can choose
any configuration for it. We have used several different initial configurations for $\vec{A}$; for all the
results presented here, we have used two choices. The first one is defined by the condition $\vec{C}=0
\Rightarrow \vec{A} = \tfrac{\hbar c}{4e}\vec{j}$, by Eqs.\ \eqref{eq:jdef} and \eqref{eq:Cdef}. This choice
enables us to test the validity of the program by comparing the energies of the initial states with those
obtained for the pure FS model. The second initial configuration has been constructed by solving Amp\`ere's
law for $\vec{A}$ with fixed $\Psi_\alpha$. Note that since we use gradient methods to minimize the energy, we
already have encoded Amp\`ere's law in our program: the gradients of the Lagrangian with respect to $A_k$
indeed yield Amp\`ere's law.

As a further test of discretization we used the same initial configurations in lattices of different sizes and
with different lattice constants. With lattice sizes of $240^3$ and above the energies were within 1\% of each
other.

\subsection{Comparison with previous FS studies}

We have also compared the new calculations with those presented in Ref.~\cite{Hietarinta:2000ci}.  By using an
initial configuration, where $\rho \equiv 1$, $\vec{C} \equiv 0$ and $V \equiv 0$, the GL model
\eqref{eq:GLfinal} reduces to the FS model \eqref{eq:trueFS}. In particular, the magnetic term becomes $
E_{M_{B}} = \tfrac{\hbar^2}{128\mu_0 e^2} \left[ \epsilon_{klm} \bigl( \vec{n} \cdot \partial_l \vec{n} \times
  \partial_m \vec{n}\bigr) \right]^2$, while in the FS model we have $ E_{T_{FS}} = \tfrac{1}{2} g_{FS}
\bigl(\vec{n} \cdot \partial_k \vec{n} \times \partial_l \vec{n} \bigr)^2.$ If we set $c=\hbar=1$,
$e=\tfrac12$, $\mu_0=1$ and $g_{FS}=\tfrac18$ (the value most commonly used in our earlier work), we have $
E_{T_{FS}} = 2E_{M_{B}}$. We created the same initial configuration with our old and new codes and found that
the energies of the initial states agree to within 1 \% with a lattice of $120^3$.

During the calculations we have monitored the topology of the system. As we saw in Sec.\
\ref{subsec:potential}, the conservation of topology is not guaranteed whenever $\rho \rightarrow 0$.
Therefore we have followed the value of the global minimum of $\rho^2$. In keeping with our earlier work, we
have also monitored the dot products of $\vec{n}\bigl(\vec{x}\bigr) \cdot
\vec{n}\bigl(\vec{x}+\vec{\mu}\bigr)$, (where $\vec{\mu} \in \{(1,0,0), (0,1,0), (0,0,1)\}$). When the global
minimum of this dot product approaches zero or becomes even negative it indicates possible breakdown of
continuity and therefore of topology during the simulation.

We have also calculated the value of the Hopf invariant directly from the field configuration.  Using the basic
differential geometric result of $H\bigl(h \circ \chi \equiv \vec{n}\bigr) = \deg \chi$, we can even use our
computational field variables. Since for any map $f \in C^\infty\bigl(\mathrm{S}^3,\mathrm{S}^3\bigr)$,
\begin{equation}\label{eq:degf}
\deg f = \tfrac{1}{2\cdot 3! \pi^2}\int_{\mathrm{S}^3}
\epsilon_{\mu\nu\sigma\rho} f^\mu \mathrm{d} f^\nu \wedge \mathrm{d} f^\sigma \wedge \mathrm{d} f^\rho,
\end{equation}
we can compute the value of the Hopf invariant by numerical integration. The Hopf invariant can also be
determined by visual inspection of the linking numbers of the preimages.

\section{\label{sec:results}Results and discussion}

A necessary condition for any stable static localized structure is its
stability against scaling. Let us assume, for now, that $\rho \ne 0$
everywhere. Then the simple scaling argument of
Derrick~\cite{Derrick:1964ww} gives a necessary condition for the
stability against scaling. Consider the behavior of the energy under
scaling $x\to\lambda x$. For the FS model in 3D the volume element of
the energy integral behaves like $\lambda^3$ while in the integrand
\eqref{eq:trueFS} the first term scales as $\lambda^{-2}$ and the
second one as $\lambda^{-4}$.  In the integral the two terms then have
opposite scaling behavior and if each is positive definite, one cannot
scale the energy to zero.  Furthermore, by the virial theorem the
preferred scaling is the one where the two terms have equal value.
In fact one can show~\cite{Vakulenko_Kapitanskii:1979} that in
the FS model the energy is bounded from below by the Hopf charge: $E>
c |Q|^{3/4}$.  This has also been confirmed
numerically~\cite{Battye:1998pe, Battye:1998zn, Hietarinta:1998kt,
  Hietarinta:2000ci}.

In the present model the scaling stability hinges on the behavior of the first
and last terms in Eq.\ \eqref{eq:GLfinal}. These terms are positive
definite and their scaling properties are opposite and thus, as long
as they are both non-zero, the system is stable. The last term,
however, equals the term $\vec{B}^2$ in Eq.\ \eqref{eq:GL} and could,
in principle, vanish. Indeed, if the $\vec C$ field obtained such a
value that
\begin{align}\label{eq:Czerocond}
  \epsilon_{klm} \partial_l C_m &= -\epsilon_{klm} \vec{n}
   \cdot \partial_l \vec{n} \times \partial_m \vec{n},
\end{align}
then the last term would vanish, rendering the system unstable against
scaling.  One could say that topological charge leaks from $\vec n$ to
$\vec C$, so that $\vec B\to 0$

We have observed the above scaling instability in direct minimizations
of all fields. The minimization does not exactly follow the route of
a uniform scaling, but we have observed that the torus-like un-knot
shrinks into a thin tube, and the value of the integrand in Eq.\
\eqref{eq:degf}, which contributes to the topological charge,
concentrates tightly in a region near the vortex core. Eventually the nearest
neighbor values in the corresponding $\vec n$ are anti-parallel, which
in effect means that the vortex configuration has shrunk to a
size less than the lattice spacing. This is accompanied with $\rho$
approaching zero, and we have verified that $\rho$ approaches zero at the same location where the neighboring
$\vec n$'s are anti-parallel.  At a place where $\rho=0$ the field $\vec n$ can no
longer be determined from $\Psi$, and the topological structure
breaks, after which the system rapidly goes to a vacuum state. We have
observed this scenario for various different potential forms and
strengths.

\begin{figure}
  \unitlength1cm
  \begin{center}
    \includegraphics[width=7cm]{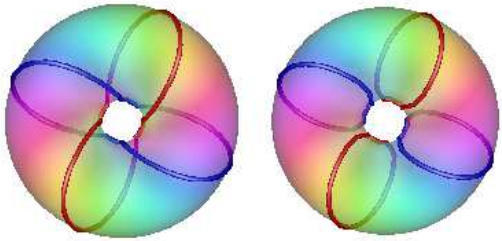}
    \caption{\label{fig:preimages}Snapshots of the preimages in the relaxation of the $(p,q) = (1,2)$ un-knot,
      before (left) and just after (right) topology breakdown. The tubes correspond to the preimages of
      $n_2=\pm 1$. The coloring on the $n_3=0$ isosurface corresponds to the longitude of $\vec n$. (Colors
      available on-line.)}
  \end{center}
\end{figure}

Two steps of the process of topology breakdown are illustrated in Fig.~\ref{fig:preimages}, where we have
plotted two distinct preimages of $\vec{n}$, $n_2=1$ and $n_2=-1$ and the isosurface $n_3=0$ on which the
tubes lie; the colours (gray-scale) of the isosurface describe longitudes. Initially (left image) the preimages
are spaced 90 degrees apart on the isosurface $n_3=0$, but during the minimization, on the region of
$xy$-plane inside the torus core, the preimages deform so that eventually there are no lattice points between
the preimages of $n_2=1$ and $n_2=-1$.  This happens even with $V_\PotRho$ (in which case $\rho>0$ everywhere)
and is not accompanied by a shrinking torus; the torus deforms somewhat but the radius of the inner
intersection of $n_3=0$ and $xy$-plane stays approximately constant. The two preimages eventually touch each
other and when they drift apart again (right image), they have become reconnected so that the topology is
trivial. After this, there is nothing to prevent, the energy from dropping to zero.

It should be noted, however, that for any non-trivial configuration
$\Psi_\alpha$, Eq. \eqref{eq:Czerocond} violates Amp\`ere's law, Eq.
\eqref{eq:Ampere}, because by using Eqs.  \eqref{eq:rotjii},
\eqref{eq:BbyCj} and \eqref{eq:Ampere}, Eq.  \eqref{eq:Czerocond}
implies $\vec{B}=0$ and $\nabla \times \vec{B} = \vec{J} = 0$, which
contradicts the topological non-triviality of $\Psi_\alpha$.  Thus any
non-trivial configuration, where Eq.  \eqref{eq:Czerocond} holds, is
unphysical. It can then be argued, that although the scaling route
leads to instability, it does so via non-physical states and thus does not
imply the instability of physically relevant states.

In order to stay within physically relevant states during minimization, we have used the following procedure:
After every minimization step applied to the $\Psi$ fields we check whether Amp\`ere's law \eqref{eq:Ampere}
is satisfied. At the beginning of a simulation the $\Psi$ fields are generated from analytical formulae, such
as \eqref{eq:Psiwithsech}, which provide the proper topological charge.  Then Amp\`ere's law is solved for
this initial state (using conjugate gradients) until is it ``sufficiently accurate''. After this initial step
we take conjugate gradient iteration steps for all fields until the convergence criterion for $\vec A$ is no
longer satisfied, at which point we continue only with $\vec A$ and Amp\`ere's law until it is satisfied to
the desired accuracy. This is repeated until the whole iteration has converged. There are various methods to
determine when the solution is ``sufficiently accurate''. Since we use a gradient-based method to solve
Amp\`ere's law, it is natural to determine the accuracy of the solution using the gradients. To accomplish
this, we observe the absolute values of the gradients affecting $\vec A$ and note the maximum of these values
at each iteration and the initial state before any iterations are made. We then compare these maxima
with the maximum found at the initial state; the solution is considered ``sufficiently accurate'' if the ratio
of the current maximum to the initial maximum is below $0.0001$.

\begin{figure*}[p]
  \unitlength1cm
  \begin{center}
    \subfigure[\mbox{}]{\includegraphics[width=8cm]{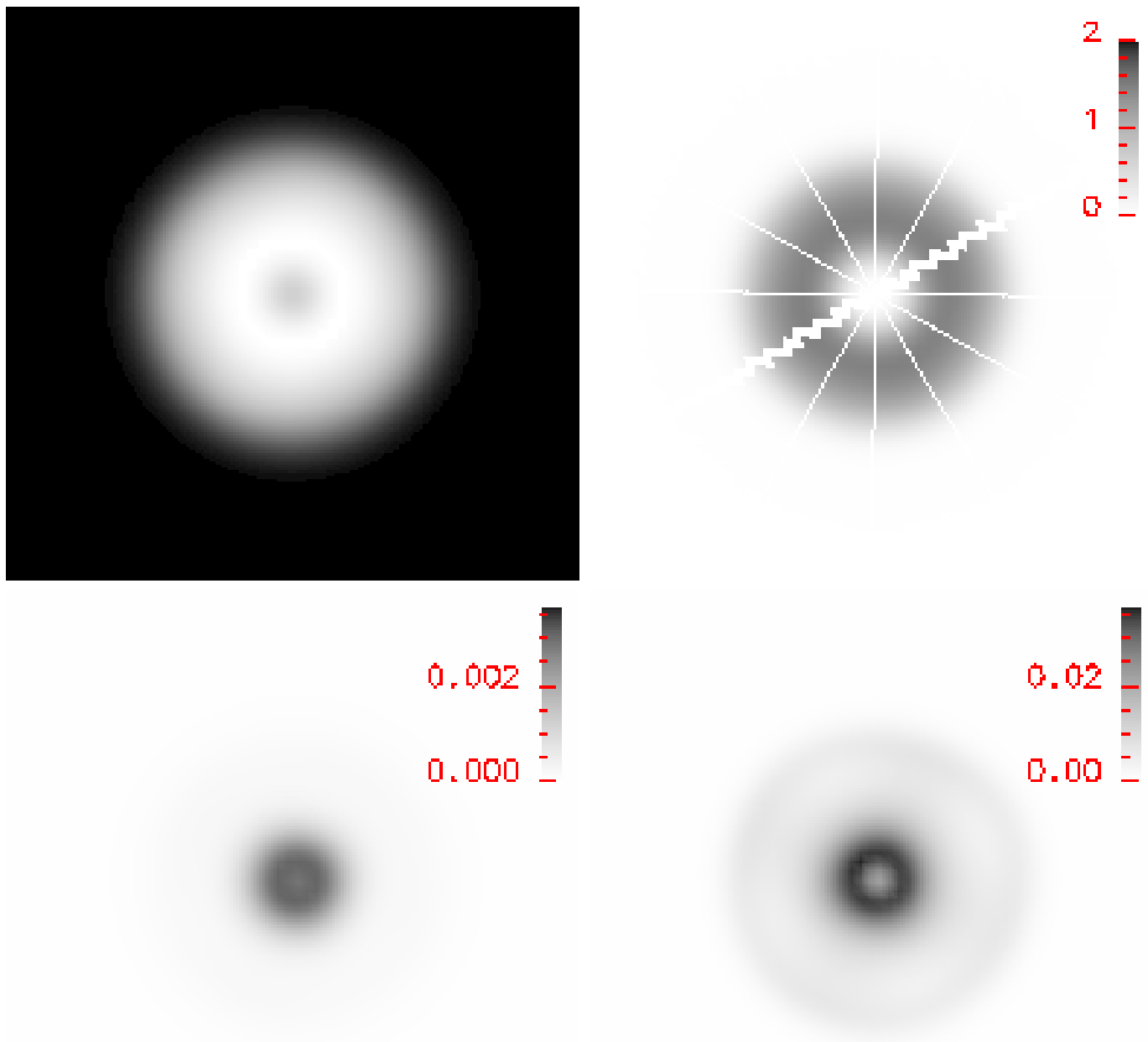}\label{fig:xy1}}
    \subfigure[\mbox{}]{\includegraphics[width=8cm]{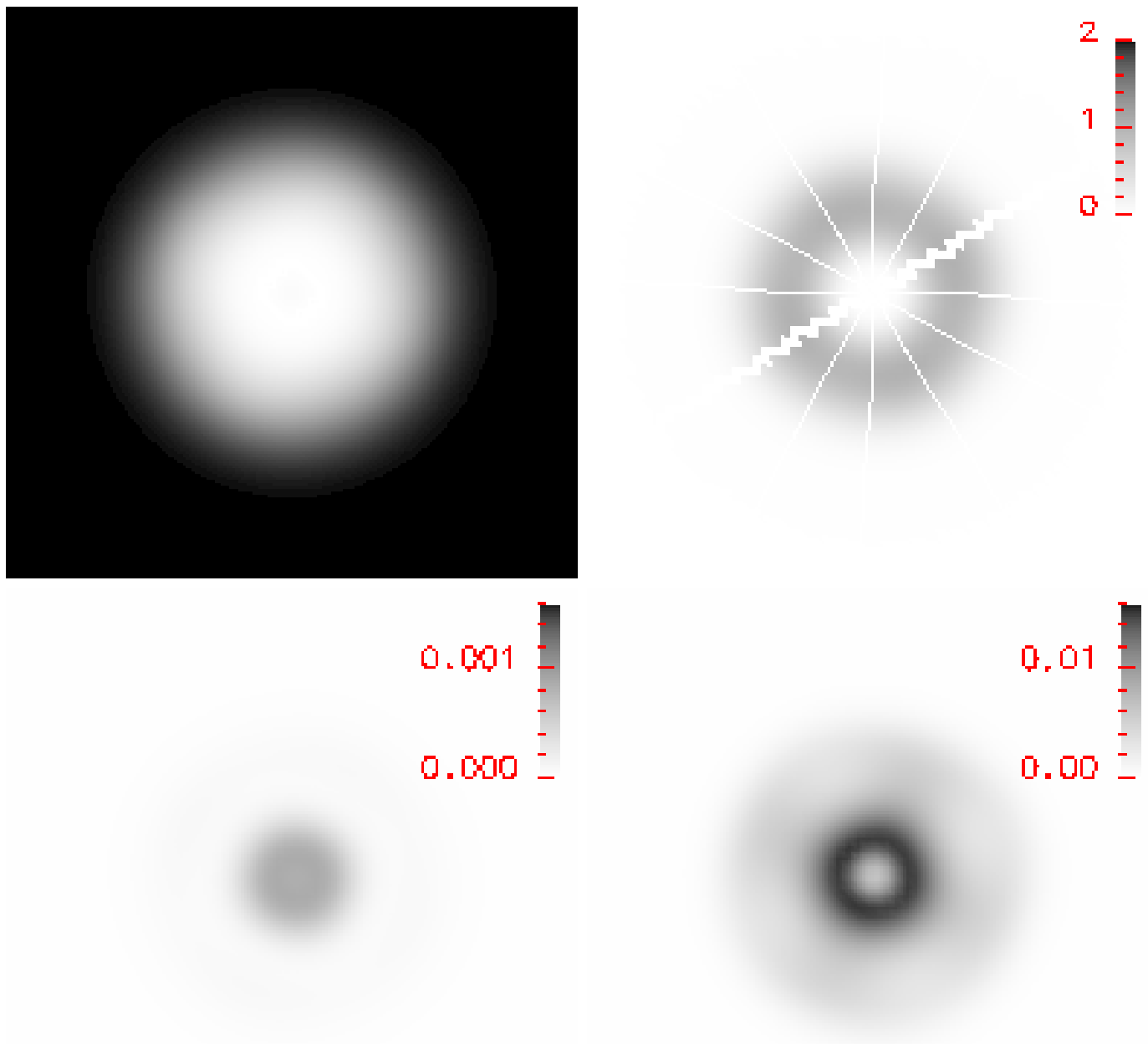}\label{fig:xy2}}
    \subfigure[\mbox{}]{\includegraphics[width=8cm]{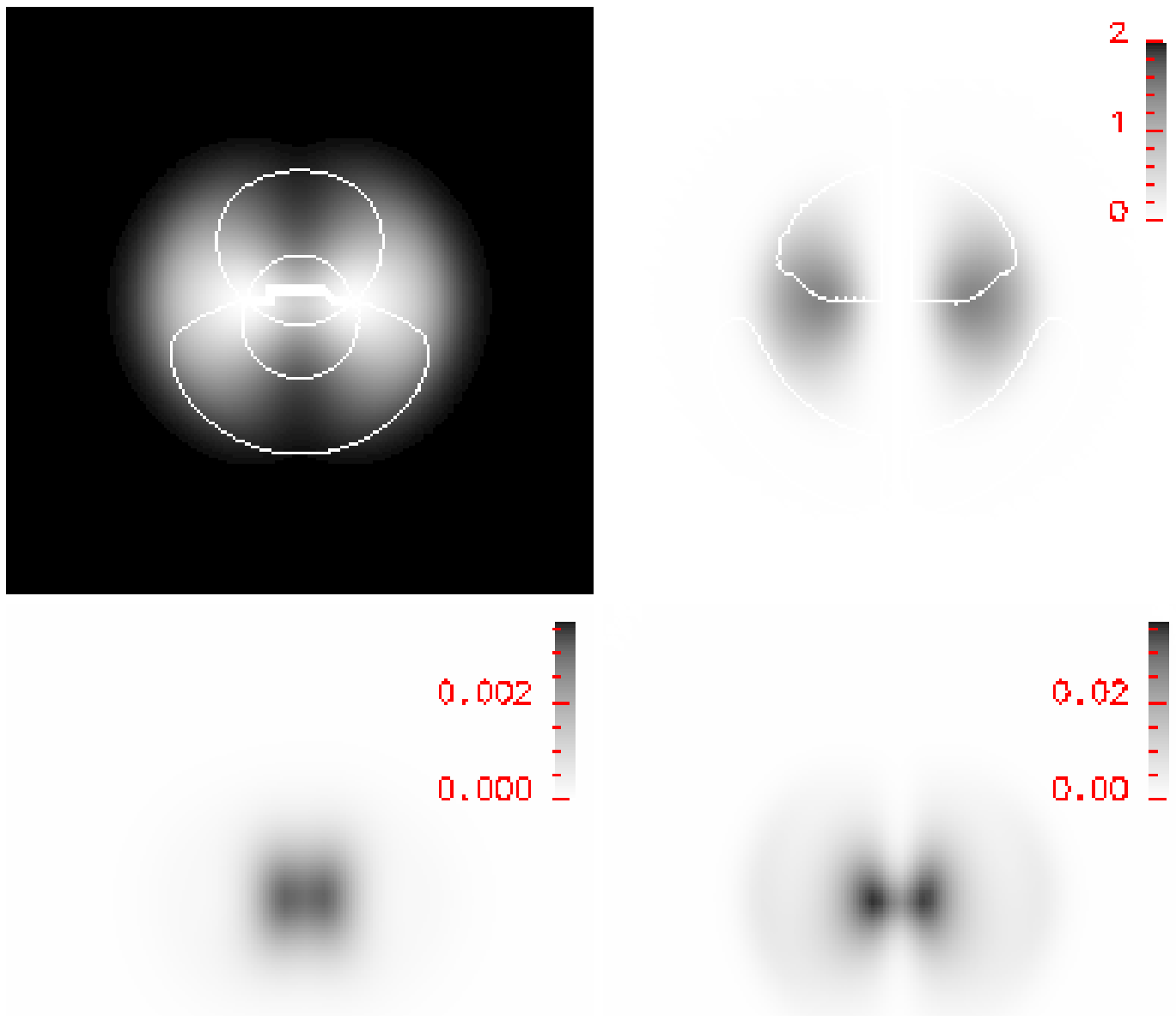}\label{fig:xz1}}
    \subfigure[\mbox{}]{\includegraphics[width=8cm]{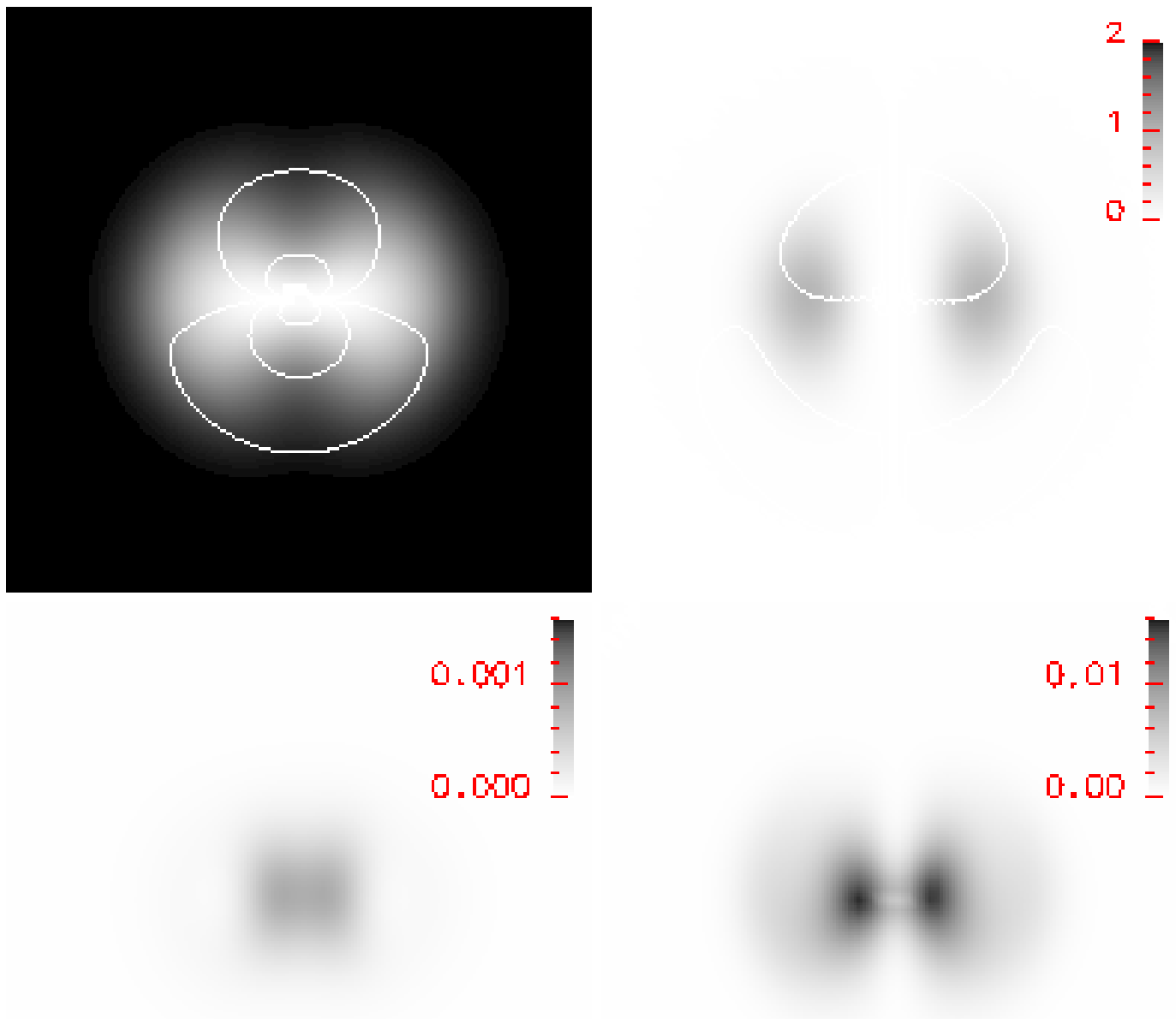}\label{fig:xz2}}
    \caption{\label{fig:densities_xy}Snapshots of a $Q=2$ system before ((a) and (c)) and just after ((b) and (d))
      the topology breakdown. Cross-sections of the system have been taken in $xy$-plane ((a) and (b)) and
      $xz$-plane ((c) and (d)). In all four panels, the upper two plots describe the densities $|\Psi_1|^2$
      (left) and $|\Psi_2|^2$ (right) on which the phases of the condensates are superimposed as contours
      spaced $\pi/3$ apart; the lower two plots of each panel describe the energy densities
      $\tfrac{1}{2}g_f|\vec{B}|^2$ (left) and $E_{\rm total}-\tfrac{1}{2}g_f|\vec{B}|^2$ (right). The jagged
      line in the $xy$-cross-sections corresponds to $\phi = \pi$; the jaggedness itself is an imaging
      artifact caused by the discreteness of the data used to produce the image.}
  \end{center}
\end{figure*}

The result of the minimization process is always the same: a singular configuration of $\Psi_\alpha$, where
the region contributing to the topological charge has shrunk to a singular line. At the same time $\vec B$ field
is generated in the hole of the torus. This can be seen from Fig.~\ref{fig:densities_xy}, which describes
states before and just after topology breakdown. In particular one can see how the $\vec B$ is formed mostly
on a smaller ring in the torus hole and how the initial maximum of $\Psi_1$ at origin vanishes while its
toroidal minimum becomes disk shaped. The topology is broken in exactly the same way as in the unconstrained
case.

The exact type of singularity and the process which leads to its
formation, depends on the potential. For $V_\PotGeneral$ and $V_\PotRho$, the small radius of
any initial torus-type isosurface of $n_3$ shrinks without limit,
snaring the region of topological interest into a singular loop. For
$V_\PotRotBab$, the process is the same unless the potential is very strong, in
which case the region where $n_2 \ne 0$, shrinks to a surface.

This process is ultimately a scaling instability, albeit a different
one from the one considered at the beginning of this Section. Now, the
scaling is not global, but only shrink the smaller circle of the
torus. The terms involving derivatives of $\Psi_\alpha$ grow without
limit during the shrinking process, but only on a loop or a surface,
allowing still the total energy to decrease.

In every case the shrinking continues until the discreteness of the computational lattice eventually breaks
the topology. 
We have tried to follow the development of this singularity by increasing lattice density. In each case a
singularity is reached, but in principle the shrinking can stop at some still smaller scale determined by the
dimensional parameters of the system (such as penetration length, strength of the potential, etc.).

\section{\label{sec:conclusions}Conclusions}

We have investigated numerically the two-component Ginzburg-Landau
model using as initial states torus un-knots, with non-vanishing Hopf change $Q$; this brings a topological structure into the system.
We have used different types of potentials depending on both order
parameters $\Psi_\alpha$ of the system.  In all cases the initial
torus tube shrinks into a thin loop and becomes untrackable in our
computational discrete lattice: the discreteness allows the
topological structure of the system to disappear eventually, contrary
to the case of the Faddeev-Skyrme model. The topological stability in
the FS model is due to the fact that the kinetic and topological terms
are non-vanishing and have an opposite behavior in the scaling. In the
GL model, the term corresponding to the FS topological term is the
magnetic field term which contributes to the unstability of the system.
In a direct minimization of the fields, the topological charge leaks from
$\vec n$ into $\vec C$ allowing eventually Derrick-type instability,
while in minimizations respecting Amp\`ere's law the process is
milder, but nevertheless leads to singularity.

Thus, it seems that the two-component GL model does not support stable topological structures having a
non-trivial conserved Hopf invariant, due to this scaling instability. It is still possible that in the
GL-system the natural size of a stable un-knot is much smaller than in the FS-model, or that it is only stable
for suitably strong potentials. It is also possible to stabilize the topological structures by adding a
suitable term in the Lagrangian. One such term has been introduced in Ref.\cite{Ward:2002vq}. These are questions
that we will study further.

\begin{acknowledgments}
  We greatly acknowledge generous computing resources from the M-grid project, supported by the Academy of
  Finland, and from CSC -- Scientific Computing Ltd., Espoo, Finland.  This work has partly been supported by
  the Academy of Finland through its Center of Excellence program.  One of us (J.J.) also wishes to thank
  Jenny and Antti Wihuri Foundation for a supporting grant.
\end{acknowledgments}

\end{document}